\documentstyle[prb,epsfig,aps,multicol]{revtex}
\def\LJ#1{$\rm LJ_{#1}$}
\def\etal{{\it et al.}}
\begin{document}
\title{The double-funnel energy landscape of the 38-atom Lennard-Jones cluster}
\author{Jonathan P.~K.~Doye\thanks{Address from September 1998:
University Chemical Laboratory, Lensfield Road, Cambridge CB2 1EW, UK}}
\address{FOM Institute for Atomic and Molecular Physics, 
Kruislaan 407, 1098 SJ Amsterdam, The Netherlands}
\author{Mark A.~Miller and David J.~Wales}
\address{University Chemical Laboratory, Lensfield Road, Cambridge CB2 1EW, UK}
\date{\today}
\maketitle
\begin{abstract}
The 38-atom Lennard-Jones cluster has a paradigmatic double-funnel energy landscape.
One funnel ends in the global minimum, a face-centred-cubic (fcc) truncated octahedron.
At the bottom of the other funnel is the second lowest energy minimum which is an
incomplete Mackay icosahedron. We characterize the energy landscape in two ways. 
Firstly, from a large sample of minima and transition states we construct a disconnectivity
tree showing which minima are connected below certain energy thresholds.
Secondly we compute the free energy as a function of a 
bond-order parameter.
The free energy profile has two minima, one which corresponds to the fcc
funnel and the other which at low temperature corresponds to the icosahedral funnel
and at higher temperatures to the liquid-like state.
These two approaches show that the greater width of the icosahedral funnel, and 
the greater structural similarity between the icosahedral structures
and those associated with the liquid-like state, are the cause 
of the smaller free energy barrier for entering the icosahedral funnel from the liquid-like state
and therefore of the cluster's preferential entry into this funnel on relaxation down the energy landscape.
Furthermore, the large free energy barrier 
between the fcc and icosahedral funnels, which is energetic in origin, 
causes the cluster to be trapped in one of the funnels at low temperature. 
These results explain in detail the link between the double-funnel energy landscape
and the difficulty of global optimization for this cluster.
\end{abstract}
\begin{multicols}{2}
\section{Introduction}

Understanding the relationship between the potential energy surface (PES), or energy landscape, 
and the dynamics of a complex system is a major research effort in the chemical physics community.
For example, much theoretical work has attempted to find the features of the
energy landscape which differentiate those model polypeptides that
are able to fold rapidly to a unique native structure from those that get
stuck in misfolded states.\cite{Sali94a,Sali94b,Klimov,Li96a,Melin98a}
Similarly, the answers to a whole host of questions about glasses
lie in the energy landscape. 
Why are glasses unable to reach the crystalline state? What is the cause
of the differences between `strong' and `fragile' liquids?
What processes, at a microscopic level,
are responsible for $\alpha$ and $\beta$ relaxation?

A key concept that has arisen within the protein folding community is
that of a funnel consisting of a set of downhill pathways that converge on
a single low-energy minimum.\cite{Leopold,Bryngel95}
It has been suggested that the PES's of proteins are characterized by
a single deep funnel and that this feature underlies
their ability to fold to their native state.
Indeed it is easy to design model single-funnel PES's that result in
efficient relaxation to the global minimum,
despite very large configurational spaces.\cite{JD96c,Zwanzig92,Zwanzig95}
By contrast, polypeptides that misfold are expected
to have other funnels that can act as traps.
Attempts have been made to characterize the energy landscape of proteins
in these terms through, for example,  
mapping the connections between compact states,\cite{Leopold,Cieplak98a}
disconnectivity graphs,\cite{BandK97,Levy98a} monotonic sequences\cite{Berry97} and 
free energy profiles.\cite{Onuchic95,Socci96}
However, for the most part these studies have been limited either to simplified lattice models, 
where the most natural elementary division of the PES into basins of attraction
surrounding local minima\cite{StillW84a} is problematic,
or to short polypeptides if more realistic models are used.

An intuitive picture of the energy landscape of glasses has also been proposed
in which the crystal corresponds to a very narrow funnel which is
inaccessible from the liquid.\cite{Angell95,Still95}
Most of the PES is dominated by rugged regions where there are many funnels leading to 
different amorphous structures. Therefore, the structure that the system relaxes to
depends upon its thermal history. In this picture the different
relaxation processes result from the hierarchy of barriers in the amorphous regions
of the PES.\cite{Still95} 
Although this picture is appealing, only recently has progress been made in 
relating the details of glassy behaviour to the features of the PES.\cite{Sastry98,Angelani98,Barkema98a}
This task is hampered by the complexity of the PES's for these systems; the number of minima
is huge and characterization of the PES is hampered by the slow relaxation times.

In view of the complex energy landscapes of model proteins 
and glasses, it is also useful to have systems for which it is easier 
to understand the relationship between the PES and the dynamics.
Clusters can provide just such an alternative perspective. 
For example, the complexity of the PES (in terms of the number of minima) 
can be controlled through the cluster size\cite{HoareM76,Tsai93a} 
and the choice of potential parameters, such as the range of attraction.\cite{JD96b,Miller98} 

Clusters where the atoms interact through the Lennard-Jones (LJ) potential---
\begin{equation}
E = 4\epsilon \sum_{i<j}\left[ \left(\sigma\over r_{ij}\right)^{12} - \left(\sigma\over r_{ij}\right)^{6}\right],
\end{equation}
where $\epsilon$ is the pair well depth and $2^{1/6}\sigma$ is the 
equilibrium pair separation---provide a particularly useful model system,
because their structure, thermodynamics and dynamics have been much studied. 
For small LJ clusters a complete enumeration of the minima and transition states allows 
a detailed view of the dynamics to be obtained.\cite{WalesB90a,Miller97}

\begin{figure}
\begin{center}
\epsfig{figure=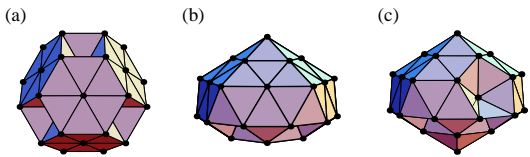,width=8.2cm}
\begin{minipage}{8.5cm}
\caption{\label{fig:piccies} 
(a) The LJ$_{38}$ global minimum, an fcc truncated octahedron ($E=-173.928427\epsilon$; point group $O_h$).
(b) Second lowest energy minimum of LJ$_{38}$ ($E=-173.252378\epsilon$; point group $C_{5v}$). 
(c) Third lowest energy minimum ($E=-173.134317\epsilon$; point group $C_{s}$). 
The structures in (b) and (c) are both incomplete Mackay icosahedra.
}
\end{minipage}
\end{center}
\end{figure}

At larger sizes, well-chosen examples allow one to consider particular 
paradigmatic types of energy landscape.
For example, the PES of \LJ{55} has a single deep funnel which leads down to the 
Mackay icosahedron\cite{Mackay} global minimum,
and \LJ{38}, the cluster which we study here, has a double-funnel landscape.
The \LJ{38} global minimum is a face-centred-cubic (fcc)
truncated octahedron\cite{JD95c,Pillardy} (Figure \ref{fig:piccies}a)
and the second lowest energy minimum is an incomplete Mackay
icosahedron\cite{Deaven96} (Figure \ref{fig:piccies}b).
These two minima lie at the bottom of separate funnels on the PES.
The thermodynamics of this cluster have recently been characterized.\cite{JD98a,JD98d}
At low temperature the truncated octahedron has the lowest free energy
but, because the entropy associated with the icosahedral funnel is larger,
its free energy 
becomes lower than that for the fcc funnel at about two thirds of the melting temperature.
Therefore, a transition takes place between the two states which is the finite-size 
equivalent of a solid-solid phase transition.

This thermodynamic transition affects the dynamics.
Relaxation down the PES from the liquid-like state almost invariably leads into 
the icosahedral funnel.
This is partly because, near to melting, icosahedral structures
have a lower free energy than fcc structures. 
However, entry into the icosahedral funnel also seems to be dynamically favoured,
perhaps because of the greater structural similarity
between the icosahedral and liquid-like structures---both have some polytetrahedral 
character.\cite{JD96c,JD96b,NelsonS}
One aim of this paper is to characterize in detail the reasons for the 
greater accessibility of the icosahedral funnel.

Furthermore, once the cluster enters the icosahedral funnel it becomes trapped
there, even when the free energy of the fcc funnel is lower.
There is a large free energy barrier between the two funnels
which prevents the cluster passing between them, but the nature 
and size of this barrier have not yet been probed.

These features make global optimization of this system much more difficult than 
for most small LJ clusters; 
most have global minima based on the Mackay icosahedra with no other 
competitive morphologies.\cite{Northby87}
Indeed, the \LJ{38} global minimum was initially discovered on the basis of 
physical insight.\cite{JD95c}
Although the truncated octahedron has since been found by a number of
global optimization methods,\cite{Pillardy,Deaven96,WalesD97,Leary97,Niesse96a,Barron96}
most of these techniques examine not the usual LJ energy landscape, but a landscape that 
has been transformed with the aim of making global optimization easier.
The basis for the success of one of these methods lies in
the significant changes to the thermodynamics 
and dynamics that the transformation causes.\cite{JD98a,JD98d}

We use two tools to characterize the double-funnel topography of the \LJ{38} PES:
in section \ref{sect:disc} we use a disconnectivity graph,
and in section \ref{sect:free} we examine the free energy landscape.
Some of the results presented here have previously appeared in a short communication.\cite{WalesMW98}

\section{Disconnectivity graph}
\label{sect:disc}

To examine the topography of the PES, we need to locate its minima and
the network of transition states and pathways that connect them.
A transition state is a stationary point on the PES where the Hessian
matrix has exactly one negative eigenvalue. Transition states
can be found efficiently
using eigenvector-following,\cite{Pancir74a,Cerjan,Wales94b} in which
the energy is maximized along one direction and simultaneously
minimized in all others. The minima connected to a transition state
are defined by the end-points of the steepest-descent paths commencing parallel and
antiparallel to the transition vector (the eigenvector whose eigenvalue
is negative). For this calculation we employ a method that uses analytic second
derivatives.\cite{Page88}
\par
The number of locally stable structures that
\LJ{38} can adopt is too large for it to be desirable or even possible to catalogue
them all. However, here we are primarily interested in the energetically
low-lying regions of the PES associated with the two funnels. In recent years, a number
of similar approaches for systematically exploring a PES by hopping
between wells have been developed,\cite{Tsai93a,JD97a,Barkema96a,Mousseau97}
and these are easily adapted to produce an algorithm that explores low-energy
regions of the PES preferentially. In our scheme, we commence at a known
low-lying minimum and proceed as follows.
\begin{enumerate}
\item Search for a transition state along the Hessian eigenvector with the smallest
non-zero eigenvalue.\label{item:search}
\item Find the steepest-descent pathway through the transition state and the
two minima it connects, as described above.\label{item:pathway}
\item There are various possible outcomes from step \ref{item:pathway}:
   \begin{enumerate}
   \item In most cases, one of the connected minima is the minimum from
which the transition state search was initiated. If this is the case, and the
other minimum is lower in energy, we move to it.
   \item If the original minimum is one of the connected ones, but the other
is higher in energy, the move is rejected.
   \item Sometimes the transition state is not connected to the minimum from
which the search started. If neither minima has been found previously,
the pathway is then isolated from the rest of the database. Since we want
to explore patterns of connectivity in the low-energy regions of the PES,
we discard the transition state and both minima under these circumstances. Such
searches can be repeated later, when the database has grown and a connection
may be found.
   \item If the original minimum is not connected, but one or both minima
have already been visited, the pathway is recorded, but we remain at the
original minimum.
\end{enumerate}
\item The procedure continues from step \ref{item:search}, searching
in both directions along
eigenvectors with successively higher eigenvalues from the modified position.
\item When a specified number, $n_{\rm ev}$, of eigenvectors of a minimum
have been searched for transition states, the position jumps to the
lowest-energy known minimum for which fewer than $n_{\rm ev}$
eigenvectors have been searched.
\end{enumerate}

By only accepting downhill moves, this algorithm prevents
the search becoming lost in the manifold of liquid-like minima. In the
present work, we chose $n_{\rm ev}=10$, allowing up to 20 transition state
searches from each minimum. In general, searches along eigenvectors with
low eigenvalues are more likely to converge to transition states in
a reasonable number of iterations. One can obtain an impression of how
thoroughly the low-energy regions of the PES have been explored by
monitoring how many of the lowest-energy minima are displaced
as the search proceeds. Having collected 3000 minima, the next 1000
displaced only 6 of the previous lowest 200, and the next 2000 displaced
only a further 7.
\par
In the course of the search, several multiple-step paths between the
low-lying $O_h$ and $C_{5v}$ minima emerged. The highest point on the
lowest-energy path that we found was a transition state with
energy $-169.709\epsilon$ ($4.219\epsilon$
above the global minimum); the path has an integrated length of $18.517\sigma$.
This demonstrates the efficiency of well-hopping techniques over more
conventional methods for exploring a PES, such as molecular dynamics (MD):
to restrict MD to regions of low potential energy, the
total energy of the simulation would have to be low, and the trajectory
would waste time undergoing intrawell oscillations, but at energies
high enough to allow interfunnel passage at a reasonable rate,
the trajectory could escape into the numerous liquid-like structures.
\par
One way to analyse a database of minima and transition states is in
terms of monotonic sequences,\cite{Berry97,BerryK95,Ball} i.e.~connected
sequences of minima with monotonically decreasing energy, which terminate
at a particular minimum. A funnel can then be defined as containing
all minima which lie on monotonic sequences to the lowest common minimum.
The primary funnel terminates at the global minimum, and is separated
from secondary funnels by primary divides---higher-lying minima lying
at the boundary between funnels. Classifying regions of the PES in this
way is useful because motion between minima in a given funnel is
likely to occur on a shorter time scale than flow between
funnels.\cite{BerryK95,Kunz95}
\par
One can also gain insight into dynamical
features of a system by examining the energy profile of the monotonic
sequences. If the decrease in energy between minima in the sequence
(i.e.~the energy gradient towards the funnel bottom) is large compared
with the intervening barriers, so that the profile looks like a staircase,
relaxation towards the funnel bottom is likely to be easy. This approach
has been used to explain the structure-seeking properties of a KCl
cluster.\cite{Ball} In contrast, energy profiles that have a shallow
gradient and high barriers are more likely to produce glass-formers.
\par
An informative way to visualize the PES and reveal funnel structure
is to plot a disconnectivity graph.\cite{BandK97}
Such graphs have been used to obtain insight into the energy landscape of
polypeptides,\cite{BandK97,Levy98a} C$_{60}$,\cite{WalesMW98} and water,\cite{WalesMW98}
sodium chloride\cite{JD98c} and Morse\cite{Miller98} clusters. 
At a given total energy, $E$, minima can be grouped into disjoint sets,
called basins, whose members are mutually accessible at that energy.
In other words, each pair of minima in a basin are connected directly or
through other minima by a path whose energy never exceeds $E$,
but would require more energy to reach a minimum in another basin.
At low energy there is just one basin---that containing the global minimum.
At successively higher energies, more basins come into play as new
minima are reached. At still higher energies, the basins coalesce as higher
barriers are overcome, until finally there is just one basin containing
all the minima (provided there are no infinite barriers).
\par
In a disconnectivity graph, the basin analysis is performed at a series
of energies, which are plotted on a vertical axis. At each energy, a
basin is represented by a node, and lines join nodes in one level to
their daughter nodes in the level below. The horizontal position of a node
has no significance, and is chosen for clarity. A funnel appears as a tall
stem with branches sprouting directly from it at a series of levels, indicating
the progressive exclusion of minima as the energy is decreased.
In the language of Ref. \onlinecite{WalesMW98}, this pattern resembles a palm tree.
Significant side-branching would indicate additional features of the
PES and leads to different patterns.\cite{WalesMW98}

\begin{figure}
\begin{center}
\epsfig{figure=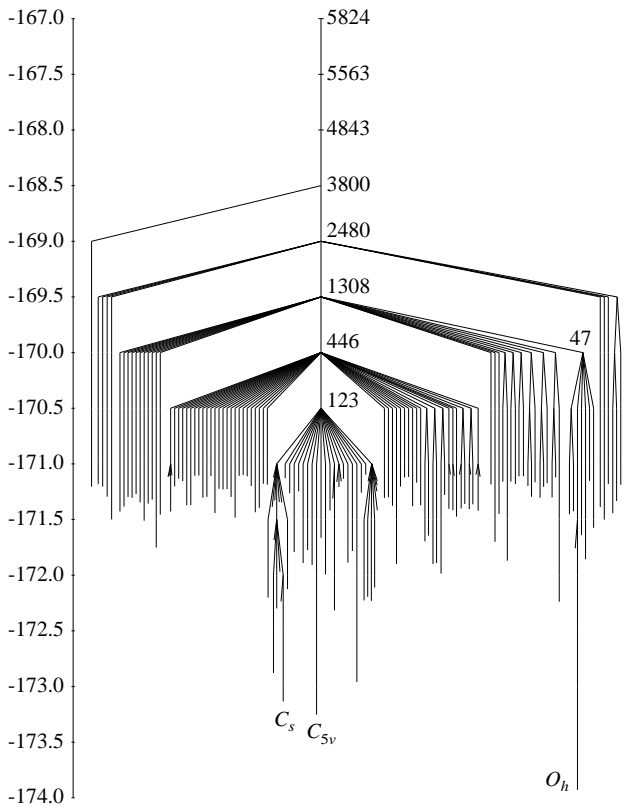,width=8.2cm}
\vglue0.2cm
\begin{minipage}{8.5cm}
\caption{\label{fig:tree}  
Disconnectivity graph for \LJ{38} using
a sample of 6000 minima and 8633 transition states. Only branches leading
to the 150 lowest-energy minima are shown, but numbers attached to nodes
indicate the number of minima they represent. The branches terminating
at the three lowest-lying minima (see Figure \ref{fig:piccies})
are labelled by
their point groups. The energy scale is in units of $\epsilon$.}
\end{minipage}
\end{center}
\end{figure}

Figure \ref{fig:tree} shows the disconnectivity graph for \LJ{38} using a sample
of 6000 minima and 8633 transition states, and a level spacing
of $0.5\epsilon$. For clarity, only branches leading
to the 150 lowest-energy minima are shown, but the number of minima that would
be represented by some of the larger nodes are indicated. The large funnel
associated with the $C_{5v}$ minimum (placed centrally) is immediately visible. 
The lowest node connecting it to the funnel of the fcc global minimum lies at
$-169.5\epsilon$, so the 446 minima in the node below can be considered as
belonging to the icosahedral funnel. In contrast, the funnel of the global
minimum contains only 47 of the minima in our sample---nearly an order of
magnitude fewer. Although only 150 branches are shown in the figure,
one can see from the rapidly increasing density of branch-ends as the
energy rises past $-171.5\epsilon$, that the number of states available to
the cluster increases dramatically with energy. This increase signifies the onset
of the liquid-like part of the PES, and the disconnectivity graph shows
that the system must enter this region of configuration space in order to pass between the
icosahedral and fcc funnels.
\par
The graph also gives an impression of the ``shape'' of the two funnels. The
branches in the fcc funnel are generally longer than those in the icosahedral
funnel, indicating higher barriers. Also, the global minimum is considerably lower in
energy than the rest of the minima in the fcc funnel because it has a complete outer shell.
In contrast to the fcc funnel, and to sizes at which a complete
Mackay icosahedron can be formed,\cite{Miller98}
the bottom of the icosahedral minimum is not dominated by a single minimum.
Rather there are many low-energy icosahedral minima associated with different 
ways of arranging the atoms in the incomplete surface layer.
It is also noticeable that the barriers about the low-lying $C_s$ minimum 
(Figure \ref{fig:piccies}c)
are lower than that for the $C_{5v}$ minimum indicating that rearrangement of 
the vacancy associated with the missing vertex atom is much easier in the $C_s$ minimum. 
These features probably explain why optimization methods found the $C_s$ minimum first;\cite{Northby87} 
the $C_{5v}$ minimum was only discovered relatively recently.\cite{Deaven96}

The overall picture, therefore, is of a narrow,
deep and somewhat rougher funnel containing the global minimum, and a broader, much
more voluminous funnel associated with the low-lying icosahedral minimum. The
``rims'' of both funnels lie in the liquid-like regions of the PES. 
The greater width of the icosahedral funnel helps to explain
why the cluster enters this region of configuration space in the vast 
majority of annealing simulations.
This effect of the funnel width has been previously observed for 
a model double-funnel PES.\cite{JD96c}

Our sample of minima is clearly only a tiny fraction
of the astronomical number available to the system, so we need to consider
the possible effects of incompleteness for the disconnectivity graph.
As discussed
above, from about half way through the search, very few new structures had
lower energy than any of the existing lowest 200. This provides good evidence
that the 150 minima actually represented in Figure \ref{fig:tree} really are
the lowest. The number of minima represented by higher-energy nodes
(for which branches have not been shown) would certainly increase if the search
were allowed to proceed for longer.
\par
The incompleteness of the transition state sample is harder to gauge, but
has important consequences for the disconnectivity graph. Given an incomplete
sample of minima, the graph depends only on transition states that interconnect
minima within the sample. Furthermore, two minima may be connected by more than
one transition state, but only the lowest matters for the graph because
it determines the energy at which the minima become mutually accessible.
When a new connectivity is discovered, the pattern of nodes and lines may change
significantly. When a lower transition state between two previously connected
minima is discovered, branching moves down the graph to lower nodes. In the
present work, up to 20 transition state searches were allowed from each minimum,
with many of the low-energy minima reaching this limit. A total of 25\,403
transition state searches were performed, 21\,515 of which converged in a
reasonable number of optimization steps. Since
only 8\,633 transition states were found, most of them must have occurred
several times, suggesting that we have an adequate sample, especially as far
as the low-energy minima are concerned.
\par
In summary, therefore, we can be quite confident that the disconnectivity graph in
Figure \ref{fig:tree} is an accurate representation of the low-energy regions
of the PES. This is because it is based on a search that is strongly biased
towards low-energy structures, and because the sample of minima and transition states
is far larger than the number of branches actually included on the graph.

\section{Free Energy landscape}
\label{sect:free}

An alternative way of characterizing the double-funnel topography of the \LJ{38} PES
is to compute the free energy as a function of a suitable order parameter.
The two funnels would be expected to give rise to minima in the free energy which
are separated by a barrier.
Bond-order parameters, which were initially introduced by Steinhardt \etal,\cite{Steinhardt83}
have been used to characterize the free energy barrier for the nucleation of
a crystal from a melt\cite{vanD92,tenW95,tenW96}
because they can differentiate between the fcc crystal and the liquid.
By calculating the bond-order parameters, $Q_4$, $Q_6$, $W_4$ and $W_6$, for our sample
of \LJ{38} minima we were able to assess whether they might also be used to differentiate
the fcc and icosahedral funnels. 
Both $Q_4$ and $Q_6$ appeared suitable and we chose to investigate $Q_4$ further.

The definition of the order parameter, $Q_l$, is
\begin{equation}
\label{eq:ql}
Q_{l} = \left(\frac{4\pi}{2l+1}\sum_{m=-l}^{l}|\overline{Q}_{lm}|^{2}\right)^{1/2}, 
\end{equation}
where 
\begin{equation}
\overline{Q}_{lm}={1\over N_b} \sum_{r_{ij}<r_0} Y_{lm}(\theta_{ij},\phi_{ij}),
\end{equation}
where the sum is over all the $N_b$ `bonds' (pairs of atoms which have a pair separation, $r_{ij}$, 
which is less than the nearest-neighbour criterion, $r_0$) in the cluster, $Y_{lm}(\theta,\phi)$ is 
a spherical harmonic and $\theta_{ij}$ and $\phi_{ij}$ are the polar and azimuthal angles of
an interatomic vector with respect to an arbitrary coordinate frame 
($Q_l$ is independent of the choice for this coordinate frame).
We use $r_0=1.391\sigma$.

\begin{figure}
\begin{center}
\epsfig{figure=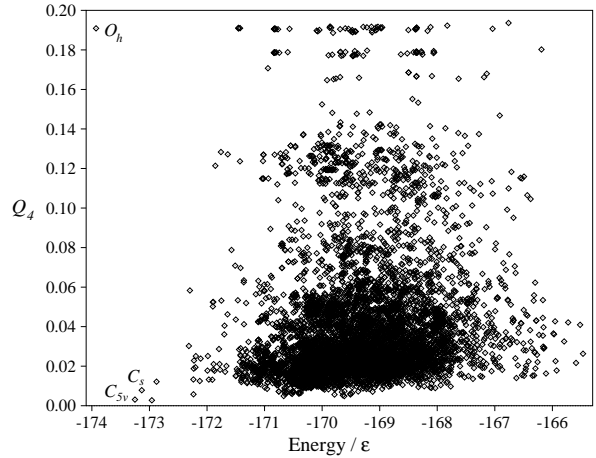,width=8.2cm}
\vglue0.1cm
\begin{minipage}{8.5cm}
\caption{\label{fig:Q4.min} $Q_4$ and potential energy values for a set of 7000 \LJ{38} minima.
The points associated with the three lowest-energy minima are labelled by their point group.}
\end{minipage}
\end{center}
\end{figure}

Figure \ref{fig:Q4.min} shows that the $Q_4$ values of the two lowest energy minima 
are well-separated. 
However, the large number of minima associated with the liquid-like state have
values only slightly greater than those for the icosahedral structures.
Therefore, $Q_4$ is a good order parameter for distinguishing fcc structures but 
not for differentiating the icosahedral and liquid-like structures.\cite{LyndenW}
The similar values of $Q_4$ for the icosahedral and liquid-like minima reflects
structural similarities---both have significant polytetrahedral character.\cite{JD96b,NelsonS}

\begin{figure}
\begin{center}
\epsfig{figure=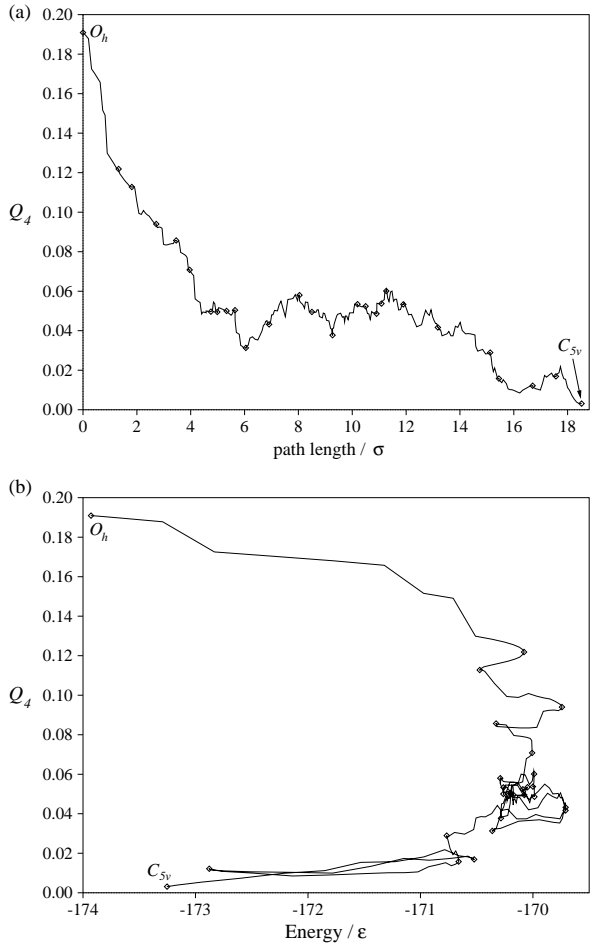,width=8.2cm}
\vglue0.1cm
\begin{minipage}{8.5cm}
\caption{\label{fig:path} (a) $Q_4$ as a function of the distance along the reaction pathway
(starting at the global minimum)
and (b) $Q_4$ against potential energy for the lowest 
energy pathway between the two lowest energy \LJ{38} minima.
The diamonds are for the stationary points (minima and transition states) on the pathway.}
\end{minipage}
\end{center}
\end{figure}

In Figure \ref{fig:path} we show the properties of the lowest-energy pathway 
between the two lowest-energy minima that we found in section \ref{sect:disc}.
The value of $Q_4$ rapidly decreases as the cluster leaves the fcc funnel 
and enters the liquid-like region of configuration space. 
A number of rearrangements take place between liquid-like minima
before the cluster then enters the icosahedral funnel.
From the pathway we can obtain upper bounds 
to the free energy barriers between the two funnels at zero temperature:
the $T=0$ free energy barriers must be less than $4.22\epsilon$ and $3.54\epsilon$
for fcc to icosahedral and icosahedral to fcc transitions, respectively.
These values are upper bounds because there may be lower-energy 
pathways that we were unable to find. 
Also the free energy difference between the two minima at $T=0$
is simply the energy difference, $0.68\epsilon$.

\begin{figure}
\begin{center}
\epsfig{figure=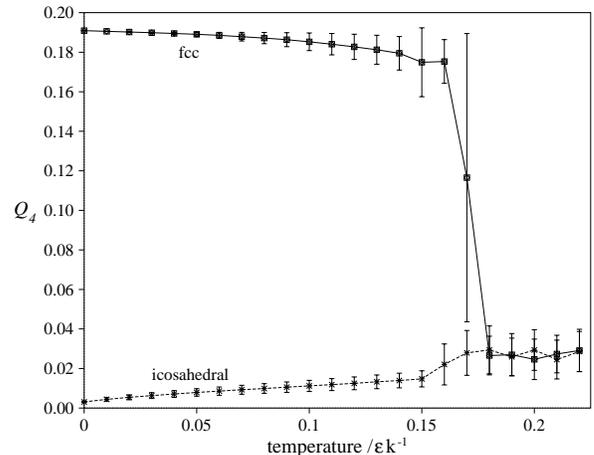,width=8.2cm}
\vglue0.1cm
\begin{minipage}{8.5cm}
\caption{\label{fig:Q4.temp} $Q_4$ for two series of canonical
Monte Carlo runs of increasing temperature. 
The initial configurations were
the fcc global minimum (solid line) and the lowest-energy icosahedral minimum (dashed line).
Each point is the average value in a $2\times 10^6$ cycle Monte Carlo run and each run was initiated from
the final geometry in the previous lower temperature run.
The error bars represent the standard deviation of the $Q_4$ probability distributions.}
\end{minipage}
\end{center}
\end{figure}

To confirm that $Q_4$ is not only able to distinguish fcc and icosahedral minima, but also
more general configurations from within the two
funnels, we performed two series of Monte Carlo simulations
of increasing temperature that started from the two lowest energy minima.
The probability distributions of $Q_4$ for the two runs are well separated and do not overlap until
until the cluster starts to melt at $T\sim 0.17\epsilon k^{-1}$ (Figure \ref{fig:Q4.temp}).
Below the melting temperature the clusters remain in the funnel 
in which the simulations were started even when the other funnel has a lower free energy. 
Hence there is a free energy barrier between the two funnels which is significantly
larger than the thermal energy.

In the canonical ensemble the Landau free energy is related to the equilibrium 
probability distribution of the order parameter by
\begin{equation}
A_L(Q_4)=A-kT\log p_{\rm can}(Q_4),
\end{equation}
where $A$ is the Helmholtz free energy.
However, conventional simulations are unable to provide equilibrium probability distributions for $Q_4$ 
because, as Figure \ref{fig:Q4.temp} illustrates, the cluster is unable to pass
over the free energy barrier between the two funnels.
To overcome this difficulty we use umbrella sampling.\cite{Torrie}
In this method configurations are not sampled with a Boltzmann distribution
but with the distribution $\exp(-\beta E+ W(Q_4))$ where $W(Q_4)$ is a biasing potential, 
the aim of which is to make configurations near the top of the free energy 
barrier more likely to be sampled.
The canonical probability distribution is then obtained from
the probability distribution for the biased run, $p_{\rm multi}(Q_4)$, by
\begin{equation}
p_{\rm can}(Q_4)=p_{\rm multi}(Q_4) \exp(-W(Q_4)).
\end{equation}
We wish to choose $W$ such that $p_{\rm multi}(Q_4)$ is approximately
constant over the whole range of $Q_4$
(the so-called multicanonical approach\cite{Berg91,Berg92}).
However, this only occurs when $W(Q_4) \approx A_L(Q_4)/kT$
and so we have to construct $W$ iteratively from the results of a number of
short preliminary simulations.\cite{Berg96}

We were able to compute $A_L(Q_4)$ successfully for $T\ge 1.5\epsilon k^{-1}$.
However, at lower temperatures, even with a reasonable biasing distribution,
the rate at which the system passed between the two funnels was too low for accurate
free energies to be obtained.
At the top of the free energy barrier in this temperature range the contribution 
of states which mediate transitions between the two funnels is small (instead the contribution of
other lower-energy states dominates), and so even at the top of the barrier a 
transition to the other funnel is an activated process. 
To overcome this difficulty would require either unfeasibly long simulations 
or a better order parameter for which the contribution of irrelevant states to the
free energy barrier region is less.
However, the results we obtain are sufficient to give us a good picture of the free 
energy landscape of \LJ{38}.

\begin{figure}
\begin{center}
\epsfig{figure=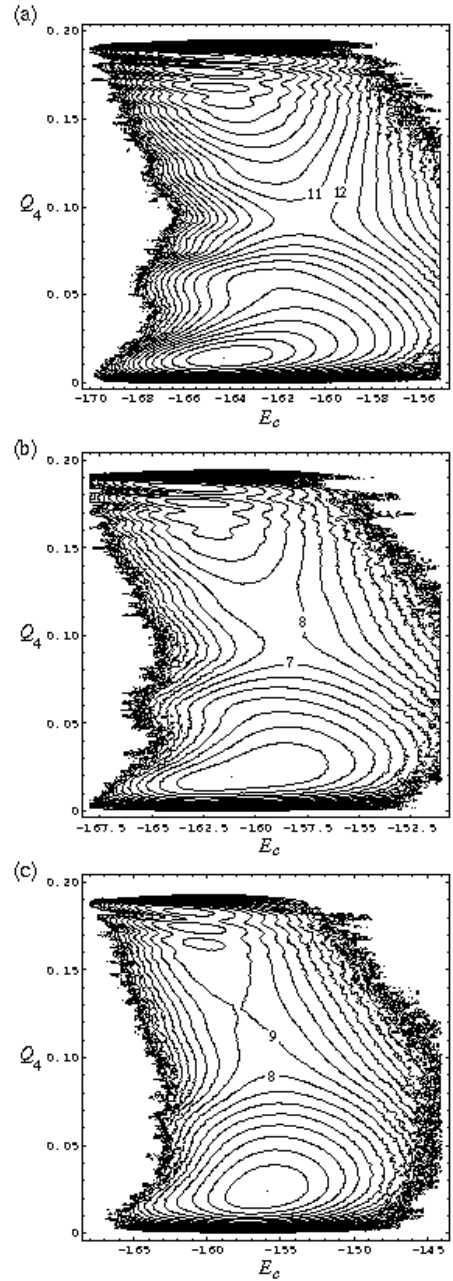,width=6.5cm}
\vglue0.1cm
\begin{minipage}{8.5cm}
\caption{\label{fig:2D} Contour plots of $A_L(Q_4,E_c)$ at 
temperatures of (a) $0.15\,\epsilon k^{-1}$, 
(b) $0.18\,\epsilon k^{-1}$ and (c) $0.21\,\epsilon k^{-1}$. 
The contours are spaced $1\,kT$ apart. The free energy zero is the bottom of the icosahedral/liquid-like
free energy minimum. Near to the free energy transition state some of the contours have been labelled by the
value of the free energy in $kT$.
No points were sampled in the regions free of contours.
The results are from MC runs of (a) $250 \times 10^6$, (b) $20 \times 10^6$ and (c) $15 \times 10^6$ cycles.}
\end{minipage}
\end{center}
\end{figure}

\begin{figure}
\begin{center}
\epsfig{figure=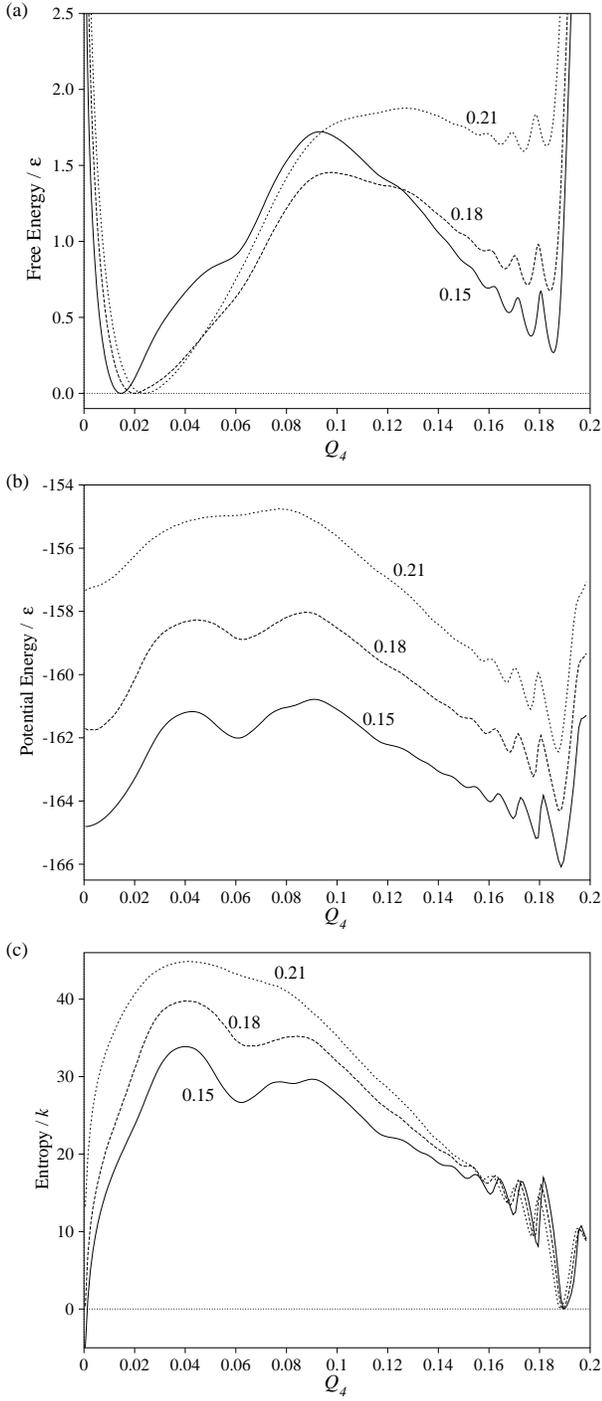,width=8.2cm}
\vglue0.1cm
\begin{minipage}{8.5cm}
\caption{\label{fig:barriers} 
(a) $A_L(Q_4)$ (b) $E_{c,L}(Q_4)$ and (c) $S_L(Q_4)$ at 
three different temperatures, as labelled (in units of $\epsilon k^{-1}$).
In each case, we have chosen the low $Q_4$ free energy minimum as the zero of $A_L$ and 
the lowest value of $S_L$ in the fcc funnel as the zero of $S_L$.
}
\end{minipage}
\end{center}
\end{figure}

In our simulations we collected the values of $Q_4$ and $E_c$ (the configurational,
or potential, energy) into a two-dimensional histogram. 
This approach allows us to obtain the two-dimensional free energy surface, 
$A_L(Q_4,E_c)$ (Figure \ref{fig:2D}), 
as well as the free energy profile, $A_L(Q_4)$ (Figure \ref{fig:barriers}a). 
Furthermore, it allows us to decompose $A_L(Q_4)$ into its energetic and entropic components,
\begin{equation}
A_L(Q_4)=E_{c,L}(Q_4)-TS_L(Q_4),
\end{equation}
because we can obtain $E_{c,L}(Q_4)$ from our two-dimensional probability distribution 
using 
\begin{equation}
E_{c,L}(Q_4)=\int p_{\rm can}(Q_4,E_c) E_c dE_c.
\end{equation}

Moreover, we can apply the histogram method\cite{McDonald} to calculate results for temperatures other than
those at which we performed simulations.
As 
\begin{equation}
p(Q_4,E_c;\beta)=\Omega_c(Q_4,E_c)\exp(-\beta E_c)/Z(\beta), 
\end{equation}
where
$\Omega_c(Q_4,E_c)$ is related to the configurational density of states by $\Omega_c(E_c)=\int\Omega_c(Q_4,E_c)dQ_4$
and $Z(\beta)$ is the partition function, it follows
that 
\begin{equation}
p(Q_4,E_c;\beta')\propto p(Q_4,E_c;\beta)\exp(-E_c(\beta'-\beta)),
\end{equation}
where $\beta$ is the reciprocal temperature of the original simulation and $\beta'$ is
the reciprocal temperature to which the results have been extrapolated.
This method, though, has to be applied with a certain amount of caution since
the extrapolation becomes increasingly sensitive to statistical errors at the edge of 
the $p(Q_4,E_c;\beta)$ distribution (Figure \ref{fig:2D}) 
as the difference in temperature increases.\cite{Newman98}

At $T=0.15\,\epsilon k^{-1}$ the free energy profile has two main minima 
(the one at $Q_4=0.015$ corresponding to the icosahedral funnel 
and the one at $Q_4=0.186$ corresponding to the fcc funnel) 
which are separated by a barrier (Figure \ref{fig:barriers}a). 
Around the fcc free energy minimum there are a number of oscillations in $A_L(Q_4)$. 
These are a result of the discontinuities that occur in the order parameter when the value of an interatomic distance
passes through $r_0$ and do not indicate that there are small free energy barriers
between different fcc structures. 
These discontinuities can also been seen in the pathway in Figure \ref{fig:path} at large values of $Q_4$.

\begin{figure}
\begin{center}
\epsfig{figure=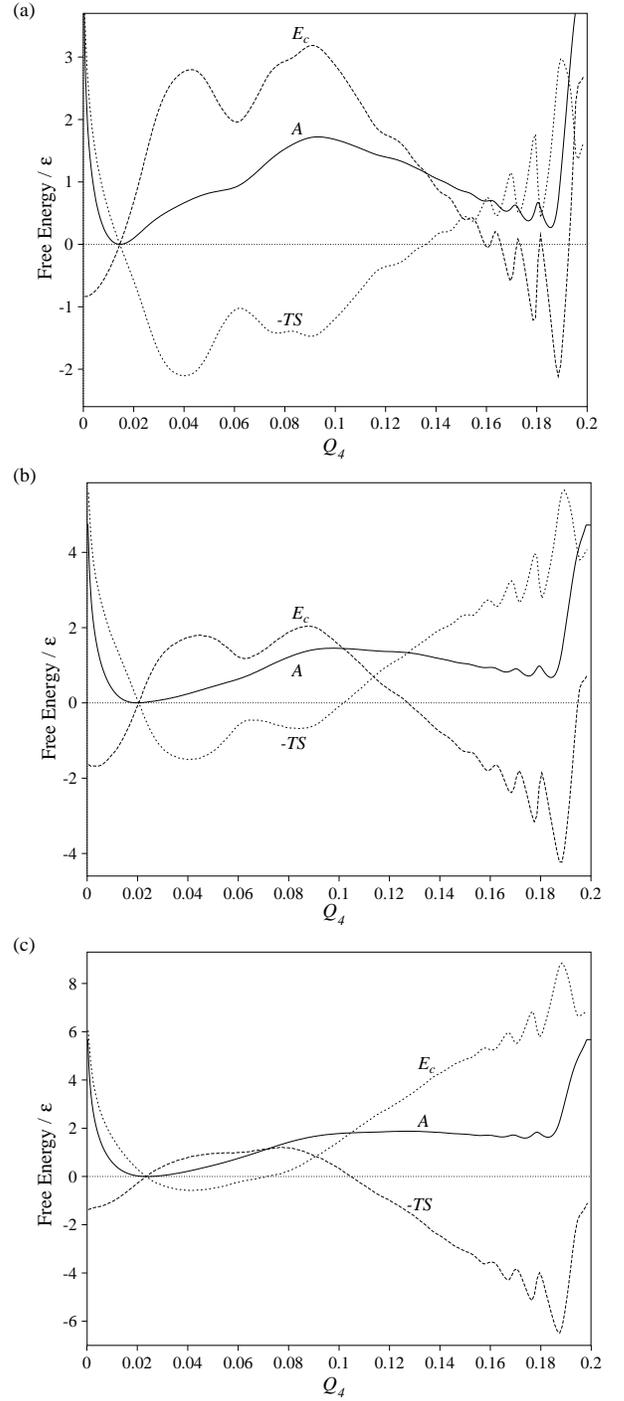,width=8.2cm}
\vglue0.1cm
\begin{minipage}{8.5cm}
\caption{\label{fig:compensate} 
The decomposition of the free energy profiles $A_L(Q_4)$ (solid line) 
into their energetic ($E_{c,L}(Q_4)$ [dashed line])  and 
entropic ($-TS_L(Q_4)$ [dotted line]) contributions at 
temperatures of (a) $0.15\,\epsilon k^{-1}$, 
(b) $0.18\,\epsilon k^{-1}$ and (c) $0.21\,\epsilon k^{-1}$. 
The zeros of all three quantities have been set to occur at the 
position of the low $Q_4$ free energy minimum.}
\end{minipage}
\end{center}
\end{figure}

\begin{figure}
\begin{center}
\epsfig{figure=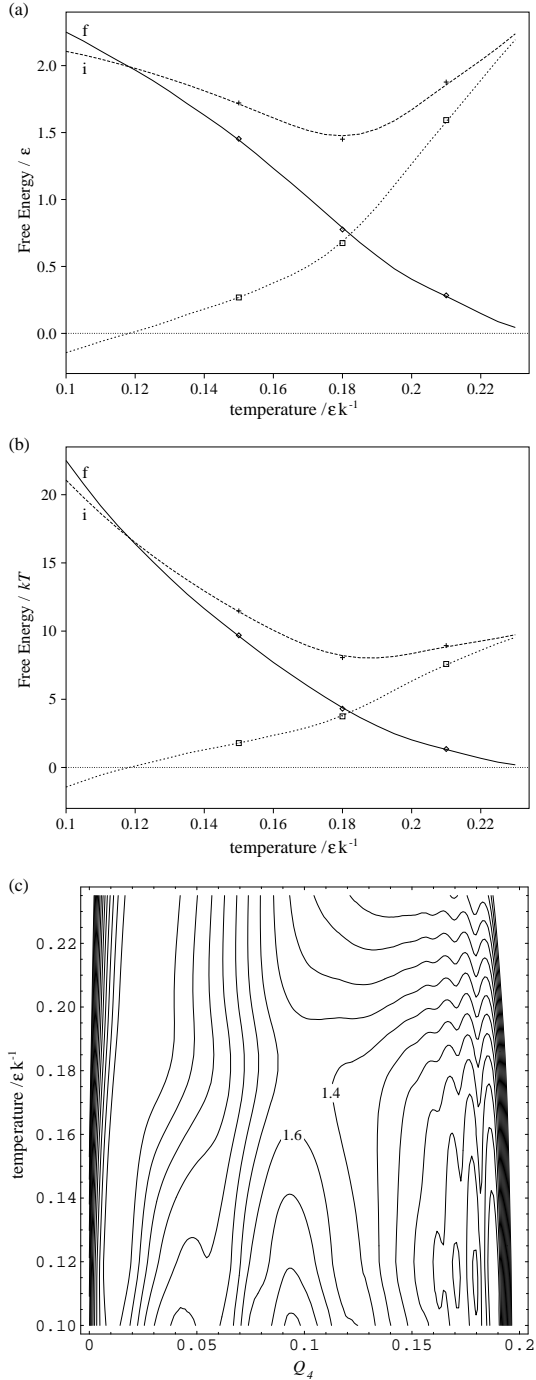,width=7.5cm}
\vglue0.1cm
\begin{minipage}{8.5cm}
\caption{\label{fig:hist} (a) and (b) The height of the free energy barrier relative to the fcc (f)
and low $Q_4$ (i) free energy minima
and the free energy difference (dotted line) between the two free energy minima as a function of temperature. 
The points are derived from the simulation profiles and the lines result from using a histogram method.
(c) Two-dimensional contour plot showing dependence of the free energy landscape, $A_L(Q_4,T)$
on temperature. 
At each temperature we have set the zero of the free energy to that of the lower free energy minimum. 
The units of free energy are $\epsilon$ in (a) and (c), and $kT$ in (b). 
The contours in (c) are at a spacing of $0.2\,\epsilon$ }
\end{minipage}
\end{center}
\end{figure}

At $T=0.15\,\epsilon k^{-1}$ the icosahedral funnel is lower in free energy because it has a larger
entropy (Figure \ref{fig:barriers}c and \ref{fig:compensate}a) due to the larger number of
icosahedral minima (Figure \ref{fig:tree}) and their lower mean vibrational frequency.\cite{JD98a,JD98d} 
The free energy barrier is large with respect to the thermal energy (11.5 and $9.69\,kT$ with respect
to the free energy minima) and relative to $kT$ increases rapidly as the temperature decreases (Figure \ref{fig:hist}b).
The size of the barrier explains why, below the melting temperature, 
simulations are trapped in one well or the other.

Examining Figure \ref{fig:compensate}a shows that at $T=0.15\,\epsilon k^{-1}$ 
the free energy barrier is energetic in origin,
and that the larger entropy of the intermediate states acts to reduce the magnitude of the barrier.
As the temperature decreases, this entropic contribution decreases and so the barrier (in absolute terms)
increases until it reaches its purely energetic value at zero temperature.
The histogram approach allows an estimate of the position of the fcc to icosahedral transition. 
It predicts that the free energy difference between the two funnels is zero at $T=0.118\,\epsilon k^{-1}$ 
(Figure \ref{fig:hist}) which is in good agreement with the thermodynamic results obtained using the superposition 
method.\cite{JD98d}

As the temperature increases from $T=0.15\,\epsilon k^{-1}$
the liquid-like state makes an increasing contribution to the 
low $Q_4$ free energy minimum.
For example, the value of $Q_4$ at the minimum gradually increases (Figures \ref{fig:barriers}a and \ref{fig:hist}c), 
reflecting the slightly larger values of $Q_4$ for the liquid-like state (Figure \ref{fig:Q4.min})
and the minimum becomes broader.
At $T=0.18\,\epsilon k^{-1}$ in a canonical simulation the cluster passes back and forth between the liquid-like
and icosahedral states in time. 
This dynamical coexistence is reflected in the low $Q_4$ free energy minimum of $A_L(Q_4,E_c)$---comparison 
of Figure \ref{fig:2D}b to Figures \ref{fig:2D}a and \ref{fig:2D}c 
shows that it is a superposition of two states.

At $T=0.21\,\epsilon k^{-1}$ the low $Q_4$ free energy minimum is solely due to the 
liquid-like state and the free energy landscape
is dominated by the free energy difference between the fcc and liquid-like structures
which results from the much larger entropy of the liquid (Figure \ref{fig:barriers}c). 
The fcc free energy minimum is now shallow and 
the flatness of the free energy landscape for $Q_4>1.0$ is a result of the compensation of the 
energy and entropic components (Figure \ref{fig:compensate}c). 
The histogram approach predicts that the fcc free energy minimum  
finally disappears at  $T\sim 0.235\epsilon k^{-1}$ (Figure \ref{fig:hist}).

\begin{figure}
\begin{center}
\epsfig{figure=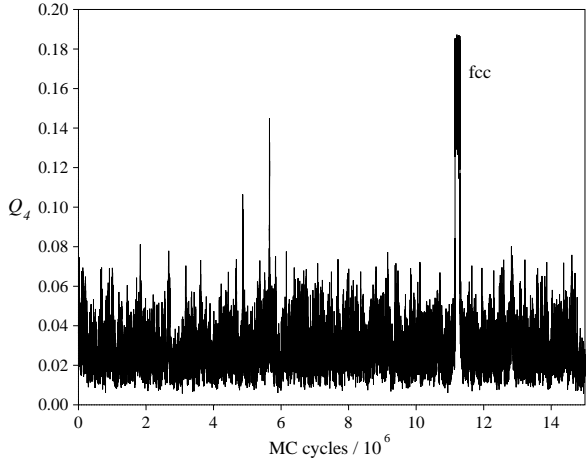,width=8.2cm}
\vglue0.1cm
\begin{minipage}{8.5cm}
\caption{\label{fig:MC} 
$Q_4$ during a canonical MC run at $T=0.19\,\epsilon k^{-1}$.
The label shows the short section of the simulation where the fcc funnel is entered.}
\end{minipage}
\end{center}
\end{figure}

The free energy barrier to pass from the low $Q_4$ free energy minimum
to the fcc funnel has an interesting temperature dependence (Figure \ref{fig:hist}). 
It has a minimum at a temperature close to the melting transition. 
Below this temperature the barrier increases because of the decreasing effect of the entropy 
of intermediate states in reducing the energetic barrier between the two free energy minima. 
Above this temperature the barrier increases because of the increasing free energy difference 
between the two minima.
As a canonical simulation is most likely to enter the fcc funnel when the free energy barrier relative
to the thermal energy is at its smallest, the optimum temperature for reaching the 
basin of attraction of the fcc global minimum is $T\sim 0.19\epsilon k^{-1}$ (Figure \ref{fig:hist}b). 
This result is somewhat counter-intuitive because one would usually expect the optimum
temperature to be when the equilibrium probability of being in the
basin of attraction of the global minimum is higher: $p^{\rm eq}_{O_h}=0.004$ 
at $T=0.19\,\epsilon k^{-1}$.\cite{JD98d} 
Figure \ref{fig:MC} does indeed confirm that a simulation at this temperature can 
enter the fcc funnel, albeit rarely.
The cluster enters the fcc funnel once in $1.5 \times 10^7$ MC cycles and
remains there for $\sim 150\,000$ cycles.

\begin{figure}
\begin{center}
\epsfig{figure=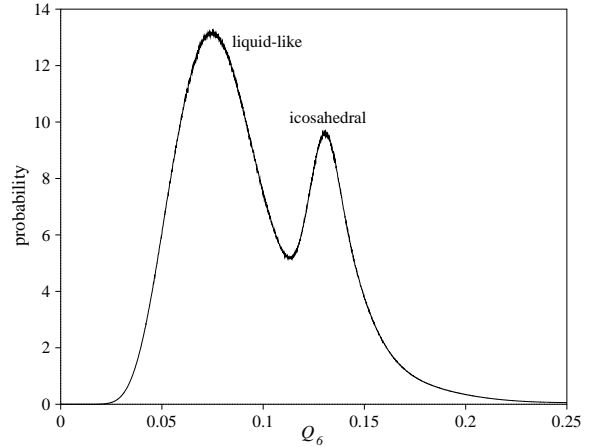,width=8.2cm}
\vglue0.1cm
\begin{minipage}{8.5cm}
\caption{\label{fig:Q6}
Probability distribution of $Q_6$ during a MC run at $T=0.18\,\epsilon k^{-1}$.}
\end{minipage}
\end{center}
\end{figure}

The order parameter $Q_4$ does not allow us to obtain the free energy barrier between the icosahedral and
liquid-like states. However, it must be considerably smaller than the barrier for passage from the 
liquid-like state to the fcc funnel, as dynamical coexistence of the two states is seen over a wide range of 
temperature. 
For \LJ{55} $E_c$ is able to detect a free energy barrier between the Mackay icosahedron and the liquid,\cite{LyndenW}
but for \LJ{38} $p(E_c)$ is unimodal.
However, $p(Q_6)$ has distinct maxima corresponding to icosahedral and liquid-like states in
the region of the melting transition (Figure \ref{fig:Q6}); 
the separate maxima disappear above $T=0.19\epsilon k^{-1}$.
At $T=0.18\,\epsilon k^{-1}$ these maxima give rise to a free energy barrier of $0.93\,kT$ 
for passing from the liquid-like state into the icosahedral funnel. 
This compares to a value of $8.08\,kT$ for passing into the fcc funnel.
This difference is consistent with the picture of the energy landscape obtained from the 
disconnectivity tree that showed the fcc funnel to be much narrower (and therefore less accessible);
it probably results from the greater structural similarity between the icosahedral and liquid-like states.
This result clearly explains why the \LJ{38} cluster is much more likely to enter the icosahedral funnel 
on relaxation down the PES.

\section{Conclusions}

The disconnectivity tree and free energy profiles that we have 
computed in this paper provide an integrated picture of the 
energy landscape of \LJ{38}.
The PES has two funnels. 
The fcc funnel is deep and narrow and 
terminates at the truncated octahedral global minimum.
The icosahedral funnel is much wider and has a flatter bottom.
The large energy barrier between the two funnels gives rise to a large
free energy barrier that causes the cluster to be trapped in one of 
the funnels when the temperature is below the melting point.
Furthermore, the difference in the width of the funnels leads to a much 
lower free energy barrier for 
passage from the liquid-like state into the icosahedral funnel. 
This explains why the cluster preferentially enters the icosahedral funnel on cooling
and why global optimization of this cluster is difficult using those methods
that do not transform the energy landscape.

\LJ{38} has a paradigmatic double-funnel energy landscape and so our understanding of this cluster
can help provide insight into other systems for which a number of funnels may be present on the PES.
There is a clear relationship between multiple funnels and trapping,
and we saw for \LJ{38} that the free energy barriers between low-energy structures in different funnels 
increase relative to the thermal energy as the temperature decreases. 
In glass-forming systems we expect that the glass transition is the result of similar
increases in free energy barriers between different low-energy amorphous structures.
It would therefore be very interesting to use disconnectivity trees to probe 
the multiple-funnel structure of the energy landscapes of model glasses.
As water is a `strong' liquid the disconnectivity tree that has been obtained for 
(H$_2$O)$_{20}$ may provide a clue to what one might expect.\cite{WalesMW98}

The trapping that results from multiple funnels on the energy landscape also 
clearly shows why it is important that proteins have a single accessible funnel.\cite{Bryngel95}
Trapping in a non-native funnel would prevent the protein from performing its biological function.
Our results for \LJ{38} also suggest the possibility that the native state might not 
always correspond to the free energy global minimum. There could be some cases where the 
funnel leading down to the global minimum is so narrow that it is inaccessible on
most time scales.\cite{JD96c} The folding of the protein would then be associated with trapping
in a more accessible funnel, and the functional lifetime of the protein would depend on the
rate of escape from this funnel to the global minimum.  
Indeed, there are some proteins from the serpin family of protease inhibitors for 
which this seems to occur.\cite{Baker}
For example, the protein plasminogen activator inhibitor-1 first folds to the active state,
but then on a time scale of hours can spontaneously transform to an inactive latent form.\cite{Franke,Carrell}
However, this scenario is likely to be the exception rather than the rule.

We found that the ease of reaching the global minimum on a multiple-funnel energy landscape 
is related to the free energy barriers for entering the different funnels from the liquid, 
which in turn are related to the width of the funnels.
Our results, therefore, add weight to the intuitive picture that 
glass-formers have a narrow funnel leading down to the crystal. 
These features of the energy landscape are intimately related to the structure.
For example, in \LJ{38} the greater accessibility of the liquid-like state can also be 
explained in terms of the greater structural similarity of the liquid to the icosahedral
structures (both have some polytetrahedral character) than to the fcc structures.
Similarly, Straley found that crystallization to a bulk close-packed structure in flat space is 
much more difficult than to the completely polytetrahedral $\{3,3,5\}$ polytope in a positively-curved space 
(the three-dimensional hypersurface of a four-dimensional sphere) where this structure 
is the global minimum.\cite{Straley84}

\acknowledgements

D.J.W.\ is grateful to the Royal Society 
and M.A.M.\ to the Engineering and Physical Sciences Research Council 
for financial support.
The work of the FOM Institute is part of the
scientific program of FOM and is supported by the Nederlandse
Organisatie voor Wetenschappelijk Onderzoek (NWO).

\end{multicols}

\begin{thebibliography}{10}

\bibitem{Sali94a}
A. Sali, E. Shakhnovich, and M. Karplus, Nature {\bf 369},  248  (1994).

\bibitem{Sali94b}
A. Sali, E. Shakhnovich, and M. Karplus, J. Mol. Biol. {\bf 235},  1614
  (1994).

\bibitem{Klimov}
D.~K. Klimov and D. Thirumalai, Phys. Rev. Lett. {\bf 76},  4070  (1996).

\bibitem{Li96a}
H. Li, R. Helling, C. Tang, and N.~S. Wingreen, Science {\bf 273},  666
  (1996).

\bibitem{Melin98a}
R. M\'elin, H. Li, N.~S. Wingreen, and C. Tang,   (cond-mat/9806197).

\bibitem{Leopold}
P.~E. Leopold, M. Montal, and J.~N. Onuchic, Proc. Natl. Acad. Sci. USA {\bf
  89},  8271  (1992).

\bibitem{Bryngel95}
J.~D. Bryngelson, J.~N. Onuchic, N.~D. Socci, and P.~G. Wolynes, Proteins:
  Structure, Function and Genetics {\bf 21},  167  (1995).

\bibitem{JD96c}
J.~P.~K. Doye and D.~J. Wales, J. Chem. Phys. {\bf 105},  8428  (1996).

\bibitem{Zwanzig92}
R. Zwanzig, A. Szabo, and B. Bagchi, Proc. Natl. Acad. Sci. USA {\bf 89},  20
  (1992).

\bibitem{Zwanzig95}
R. Zwanzig, Proc. Natl. Acad. Sci. USA {\bf 92},  9801  (1995).

\bibitem{Cieplak98a}
M. Cieplak, M. Henkel, J. Karbowski, and J.~R. Banavar, Phys. Rev. Lett. {\bf
  80},  3654  (1998).

\bibitem{BandK97}
O.~M. Becker and M. Karplus, J. Chem. Phys. {\bf 106},  1495  (1997).

\bibitem{Levy98a}
Y. Levy and O.~M. Becker, Phys. Rev. Lett. {\bf 81},  1126  (1998).

\bibitem{Berry97}
R.~S. Berry, N. Elmaci, J.~P. Rose, and B. Vekhter, Proc. Natl Acad. Sci. USA
  {\bf 94},  9520  (1997).

\bibitem{Onuchic95}
J.~N. Onuchic, P.~G. Wolynes, Z. Luthey-Schulten, and N.~D. Socci, Proc. Natl.
  Acad. Sci. USA {\bf 92},  3626  (1995).

\bibitem{Socci96}
N.~D. Socci, J.~N. Onuchic, and P.~G. Wolynes, J. Chem. Phys. {\bf 104},  5860
  (1996).

\bibitem{StillW84a}
F.~H. Stillinger and T.~A. Weber, Science {\bf 225},  983  (1984).

\bibitem{Angell95}
C.~A. Angell, Science {\bf 267},  1924  (1995).

\bibitem{Still95}
F.~H. Stillinger, Science {\bf 267},  1935  (1995).

\bibitem{Sastry98}
S. Sastry, P.~G. Debenedetti, and F.~H. Stillinger, Nature {\bf 393},  554
  (1998).

\bibitem{Angelani98}
L. Angelani, G. Parisi, G. Ruocco, and G. Viliani,   (cond-mat/9803165).

\bibitem{Barkema98a}
G.~T. Barkema and N. Mousseau, Phys. Rev. Lett.  in press  (cond-mat/9804317).

\bibitem{HoareM76}
M.~R. Hoare and J. McInnes, J. Chem. Soc., Faraday Discuss. {\bf 61},  12
  (1976).

\bibitem{Tsai93a}
C.~J. Tsai and K.~D. Jordan, J. Phys. Chem. {\bf 97},  11227  (1993).

\bibitem{JD96b}
J.~P.~K. Doye and D.~J. Wales, J. Phys. B {\bf 29},  4859  (1996).

\bibitem{Miller98}
M.~A. Miller, J.~P.~K. Doye, and D.~J. Wales, J. Chem. Phys.  submitted
  (cond-mat/9808080).

\bibitem{WalesB90a}
D.~J. Wales and R.~S. Berry, J. Chem. Phys. {\bf 92},  4283  (1990).

\bibitem{Miller97}
M.~A. Miller and D.~J. Wales, J. Chem. Phys. {\bf 107},  8568  (1997).

\bibitem{Mackay}
A.~L. Mackay, Acta Cryst. {\bf 15},  916  (1962).

\bibitem{JD95c}
J.~P.~K. Doye, D.~J. Wales, and R.~S. Berry, J. Chem. Phys. {\bf 103},  4234
  (1995).

\bibitem{Pillardy}
J. Pillardy and L. Piela, J. Phys. Chem. {\bf 99},  11805  (1995).

\bibitem{Deaven96}
D.~M. Deaven, N. Tit, J.~R. Morris, and K.~M. Ho, Chem. Phys. Lett. {\bf 256},
  195  (1996).

\bibitem{JD98a}
J.~P.~K. Doye and D.~J. Wales, Phys. Rev. Lett. {\bf 80},  1357  (1998).

\bibitem{JD98d}
J.~P.~K. Doye, D.~J. Wales, and M.~A. Miller, J. Chem. Phys.  in press
  (cond-mat/9806020).

\bibitem{NelsonS}
D.~R. Nelson and F. Spaepen, Solid State Phys. {\bf 42},  1  (1989).

\bibitem{Northby87}
J.~A. Northby, J. Chem. Phys. {\bf 87},  6166  (1987).

\bibitem{WalesD97}
D.~J. Wales and J.~P.~K. Doye, J. Phys. Chem. A {\bf 101},  5111  (1997).

\bibitem{Leary97}
R.~H. Leary, J. Global Optimization {\bf 11},  35  (1997).

\bibitem{Niesse96a}
J.~A. Niesse and H.~R. Mayne, J. Chem. Phys. {\bf 105},  4700  (1996).

\bibitem{Barron96}
C. Barr\'on, S. G\'omez, and D. Romero, Appl. Math. Lett. {\bf 9},  75  (1996).

\bibitem{WalesMW98}
D.~J. Wales, M.~A. Miller, and T.~R. Walsh, Nature {\bf 394},  758  (1998).

\bibitem{Pancir74a}
J. Panc{\'\i}\v{r}, Coll. Czech. Chem. Comm. {\bf 40},  1112  (1974).

\bibitem{Cerjan}
C.~J. Cerjan and W.~H. Miller, J. Chem. Phys. {\bf 75},  2800  (1981).

\bibitem{Wales94b}
D.~J. Wales, J. Chem. Phys. {\bf 101},  3750  (1994).

\bibitem{Page88}
M. Page and J.~W. McIver, J. Chem. Phys. {\bf 88},  922  (1988).

\bibitem{JD97a}
J.~P.~K. Doye and D.~J. Wales, Z. Phys. D {\bf 40},  194  (1997).

\bibitem{Barkema96a}
G.~T. Barkema and N. Mousseau, Phys. Rev. Lett. {\bf 77},  4358  (1996).

\bibitem{Mousseau97}
N. Mousseau and G.~T. Barkema, Phys. Rev. E {\bf 57},  2419  (1998).

\bibitem{BerryK95}
R.~S. Berry and R.~E. Breitengraser-Kunz, Phys. Rev. Lett. {\bf 74},  3951
  (1995).

\bibitem{Ball}
K.~D. Ball {\it et~al.}, Science {\bf 271},  963  (1996).

\bibitem{Kunz95}
R.~E. Kunz and R.~S. Berry, J. Chem. Phys. {\bf 103},  1904  (1995).

\bibitem{JD98c}
J.~P.~K. Doye and D.~J. Wales, Phys. Rev. B  submitted  (cond-mat/9801152).

\bibitem{Steinhardt83}
P.~J. Steinhardt, D.~R. Nelson, and M. Ronchetti, Phys. Rev. B {\bf 28},  784
  (1983).

\bibitem{vanD92}
J.~S. van Duijneveldt and D. Frenkel, J. Chem. Phys. {\bf 96},  4655  (1992).

\bibitem{tenW95}
P.~R. ten Wolde, M.~J. Ruiz-Montero, and D. Frenkel, Phys. Rev. Lett. {\bf 75},
   2714  (1995).

\bibitem{tenW96}
P.~R. ten Wolde, M.~J. Ruiz-Montero, and D. Frenkel, J. Chem. Phys. {\bf 104},
  9932  (1996).

\bibitem{LyndenW}
R.~M. Lynden-Bell and D.~J. Wales, J. Chem. Phys. {\bf 101},  1460  (1994).

\bibitem{Torrie}
G.~M. Torrie and J.~P. Valleau, Chem. Phys. Lett. {\bf 28},  578  (1974).

\bibitem{Berg91}
B.~A. Berg and T. Neuhaus, Phys. Lett. B {\bf 267},  249  (1991).

\bibitem{Berg92}
B.~A. Berg and T. Neuhaus, Phys. Rev. Lett. {\bf 69},  9  (1992).

\bibitem{Berg96}
B.~A. Berg, J. Stat. Phys. {\bf 82},  323  (1996).

\bibitem{McDonald}
I.~R. McDonald and K. Singer, J. Chem. Soc., Faraday Discuss. {\bf 43},  40
  (1967).

\bibitem{Newman98}
M.~E.~J. Newman and R.~G. Palmer,   (cond-mat/9804306).

\bibitem{Baker}
D. Baker and D.~A. Agard, Biochemistry {\bf 33},  7505  (1994).

\bibitem{Franke}
G. Franke, E.~R. Hilf, and P. Borrmann, J. Chem. Phys. {\bf 98},  3496  (1993).

\bibitem{Carrell}
R.~W. Carrell, D.~L. Evans, and P.~E. Stein, Nature {\bf 353},  576  (1991).

\bibitem{Straley84}
J.~P. Straley, Phys. Rev. B {\bf 30},  6592  (1984).

\end{thebibliography}
\end{document}